\documentclass[prl,amsmath,twocolumn,showpacs]{revtex4}

\usepackage{epsfig}

\begin{document}

\title{Time transformation for random walks in the quenched trap model}
\author{S. Burov, E. Barkai }
\affiliation{ Department of Physics, Institute of Nanotechnology and Advanced Materials,  Bar Ilan University, Ramat-Gan
52900, Israel}

%
%
\begin{abstract}
Diffusion in the quenched trap model is investigated with an approach we
call weak subordination breaking. We map the problem onto Brownian
motion and  show that the operational time
is ${\cal S}_\alpha = \sum_{x=-\infty} ^\infty (n_x)^\alpha$ 
where $n_x$ is the visitation number at site $x$ . 
In the limit of zero temperature
we recover the renormalization group (RG) solution found by Monthus. Our
approach is an alternative to RG capable of dealing with any disorder strength.
\end{abstract}


\maketitle



 Random walks in disordered systems with diverging expected waiting times 
have attracted vast interest over many decades \cite{Bouchaud,Review}. 
Two 
approaches
in this field are the annealed continuous time random walk (CTRW) model
and 
the  by far more challenging quenched trap model (QTM). Starting in the seventies,
the Scher-Montroll CTRW  approach was  used
to model sub-diffusive  photo-currents in amorphous materials 
\cite{Scher}. 
 Bouchaud showed that
the trap model
is a useful tool for the description of aging phenomena
in glasses \cite{B92,MB96}. More recently
these models were used to describe non self averaging \cite{Stas}
and weak ergodicity breaking \cite{B92,Golan} which
are important for the statistical  description 
of blinking quantum dots \cite{PhysToday} and
diffusion of single molecules in living cells \cite{Burov}.   

 This manuscript presents a new approach for random walks in a fixed random
environment. 
With physical arguments \cite{Bouchaud,Machta,Alexander1} and  rigorous
mathematics \cite{Ben,BenHarosh} we know that the  QTM
in dimensions $d>2$ 
is expected to qualitatively behave like its corresponding
mean field CTRW, the latter being exact when $d\to \infty$.  
For a random walk in a quenched disordered system
intricate correlations induced by multiple visits 
to the same site make the problem non-trivial and interesting. 
For that reason RG methods \cite{Machta,Monthus}
 were used to tackle
this problem. 
With RG  Machta \cite{Machta} found the 
scaling exponents of the QTM 
and Monthus \cite{Monthus} investigated
the diffusion front 
in the limit of zero temperature (see details below).
While the RG is powerful
it  has its limitations: 
a simple
approach which predicts the  diffusion front 
is still missing.

 We provide the long sought after breakthrough in the statistical
analysis of sub-diffusion in the QTM. Our approach is based
on a novel time transformation.
It is well known that one may decompose
the CTRW process into ordinary Brownian motion and a L\'evy process, an approach
called subordination \cite{Sokolov}.
 In this scheme
normal  Brownian motion takes place in operational time $s$.
The disorder is effectively described by a L\'evy time transformation from
operational time $s$ to  laboratory time $t$ (see some details below). 
This method is not generally
suited for random walks in quenched environments since
it uses the renewal assumption. In a quenched environment 
this very strong assumption implies  
that a given lattice site is visited only once along
the path of the random walker, i.e. it neglects correlations.  
 So a new approach 
capable of dealing with 
quenched disorder is now investigated. 
We focus on the one dimensional case since then the departure from
mean field is the strongest. 
At the end
of the Letter we explain how to extend our results to other interesting cases.

{\em Quenched trap model} \cite{Bouchaud,Monthus,Bertin}. 
 We consider a random walk on a one dimensional lattice with lattice spacing
equal one. For each lattice
site  $x$ there is a quenched random variable $\tau_x$ which is the 
waiting time between jump events for a particle situated on $x$. After waiting 
for a period $\tau_x$ the particle jumps to one of its two nearest 
neighbors with equal probability. 
The particle starts on the origin $x=0$ at time $t=0$, waits for time
$\tau_0$, then with probability $1/2$ jumps to  $x=1$ (or $x=-1$), 
waits there for
$\tau_1$ (or $\tau_{-1}$) then if the particle returns to $x=0$ it waits 
for a time interval $\tau_0$ etc. 
The $\{\tau_x\}$s are positive independent identically
distributed random variables with a common probability density function
(PDF)   
\begin{equation}
\psi(\tau_x) \sim { A \over |\Gamma(-\alpha)| } (\tau_x)^{- (1 + \alpha)}  
\label{eqTR01}
\end{equation} 
for $\tau_x \to \infty$ and $0<\alpha<1$.
Hence 
the Laplace transform of the waiting time PDF  is
$\hat{\psi}(u)\sim 1 - A u^\alpha + \cdots $
when $u \to 0$. 
As well known \cite{Bouchaud} the QTM describes a random walk among
 traps whose energy depth $E>0$ is exponentially distributed
$f(E) = \exp(-E/T_g)/T_g$ where $T_g$ is a measure of the disorder.
It is easy to show that $\alpha = T/T_g$
and $A= |\Gamma(-\alpha)| \alpha$ where
$T$ is the thermal temperature. 
The goal of this paper is to find the long time behavior
of $\langle P(x,t) \rangle$
the probability of finding the particle on $x$ at time $t$
averaged over the disorder. 
For a comprehensive mathematical review of the QTM
see \cite{BenHarosh}.  

\begin{table}[ht]
\centering
\begin{tabular}{l c c c c c c }
\hline\hline
$\alpha$ & 0 & 0.2 & 0.4 & 0.6 & 0.8 & 1 \\ 
$\langle z^2 \rangle$ & 0.5 & 0.67 & 0.81 & 0.91 & 0.96 & 1 \\
\hline
\end{tabular}
\caption{Simple Brownian simulations on a lattice give $\langle z^2 \rangle$ 
which 
according to Eq. (\ref{eqSca08}) 
yield $\langle x^2 \rangle$ for the QTM.   
}
\label{table1}
\end{table}

{\em Time in the quenched trap model}
is 
$ t = \sum_{x=-\infty} ^\infty n_x \tau_x $
where $n_x$ is  
 the number of visits to lattice point $x$.
   Define the random variable 
%
$\eta = t / ({\cal S}_\alpha)^{1/\alpha} $
%
where
\begin{equation}
{\cal S}_\alpha = \sum _{x=-\infty} ^\infty (n_x)^\alpha.
\label{eqtime03}
\end{equation}
When $\alpha=1$,  ${\cal S}_{\alpha}$ is the total number of jumps
made $\sum_{x=-\infty} ^\infty n_x = s$.
In the opposite
limit $\alpha\to 0$, 
${\cal S}_0$ is the distinct number of sites
visited by the random walker which is called the span of the random
walk. 
 We now show that in the scaling limit
\begin{equation}
\mbox{ the PDF of \ }  \eta \mbox{ \ is:\ } l_{\alpha,A,1} 
\left( \eta\right),  
\label{eqtime04}
\end{equation} 
where $l_{\alpha,A,1}(\eta)$ is the one sided L\'evy PDF
whose 
Laplace $\eta \to u$ pair is $\exp(- A u^{\alpha})$. 
Namely the heavy tailed distribution of the waiting times ${\tau_x}$ 
determines the statistics of $\eta$ through the characteristic
exponent $\alpha$, while the visitation numbers $\{ n_x \}$ provide
the scaling through ${\cal S}_\alpha$.  
By definition
the Laplace $\eta \to u$ transform of the PDF of $\eta$ is 
\begin{equation}
\langle e^{ - \eta u } \rangle = \langle \exp\left[  - \sum_{i=-\infty} ^\infty {  n_i \tau_i \over \left( {\cal S}_\alpha \right)^{1/\alpha} } u \right] \rangle.
\label{eqApA01z}
\end{equation} 
We average with respect to the 
disorder, namely with respect to the independent and identically
distributed  random waiting times $\tau_x$,
and obtain
%
$\langle e^{- u \eta } \rangle = \Pi_{x=-\infty} ^\infty \hat{\psi} \left[ { n_x u \over  ({\cal S}_\alpha)^{1/\alpha} }  \right] $
%
where $\hat{\psi}(u)$ is the Laplace transform of the PDF of waiting
times $\psi(\tau_x)$. We use 
 $\hat{\psi}(u) = \exp( - A u^\alpha)\sim 1 - A u^\alpha+ \cdots$
\begin{equation} 
\langle e^{ - u \eta} \rangle = \Pi_{x=-\infty} ^\infty \exp\left[ - {A (n_x)^\alpha u ^\alpha \over {\cal S}_\alpha }\right] = e^{ -A u^\alpha}. 
\label{eqApA02ez}   
\end{equation}   
Hence the PDF of $\eta$ is a one sided L\'evy  law Eq.  
(\ref{eqtime04}).
In a longer publication  we  will complete the proof
and consider the case where $\psi(\tau_x)$ belongs to the
domain of attraction L\'evy PDFs 
(i.e. families  of PDFs satisfying $\hat{\psi}(u) \sim 1 - A u^\alpha + \cdots$).
We now invert the process fixing time $t$ to find
the PDF of ${\cal S}_\alpha$ 
\begin{equation}
n_t \left( {\cal S}_\alpha \right) = { t \over \alpha}  ({\cal S}_\alpha)^{ - 1/\alpha - 1} l_{\alpha,A,1} \left[ { t \over ({\cal S}_\alpha )^{1/\alpha} }\right]. 
\label{eqtime05}
\end{equation}
In the following we explain how to use
the operational time ${\cal S}_\alpha$ to obtain the
desired diffusion front of the QTM, in other words we explain how
we get rid of the disorder and focus only on Brownian motion. 

\begin{figure}
\begin{center}
\epsfig{figure=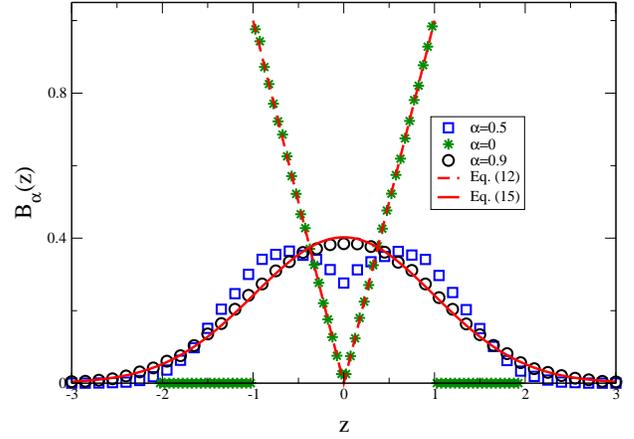,totalheight=0.34\textheight,angle=-90,
width=0.45\textwidth}
\end{center}
\caption{ The PDF  $B_\alpha (z)$ exhibits a transition between a Gaussian
shape when $\alpha \to 1$ to a $V$ shape when $\alpha\to 0$.  
Simulations of Brownian motion on a lattice yield excellent agreement
with theoretical predictions Eqs.  
(\ref{eqaplz05},
\ref{eqBa102a}) without fitting.  
}
\label{fig1}
\end{figure}

{\em Weak subordination breaking.} 
 To find
$\langle P(x,t)\rangle$ 
we follow six steps:
{\bf 1.} Choose the laboratory time $t$ which is a fixed parameter.
{\bf 2.}  Use a random number generator and draw the stable
random variable  $\eta$ from the
one sided L\'evy PDF $l_{\alpha,A,1} (\eta)$.
{\bf 3.} With $\eta$ and $t$ determine the operational time 
${\cal S}_\alpha= (t/\eta)^\alpha$.
{\bf 4.}  Generate  a simple symmetric
 random walk on a lattice (probability
$1/2$ for jumping left and right). Stop the process once its 
${\cal S}_\alpha$ reaches the operational time set in step $3$.   
{\bf 5.} Record the position $x$ of the particle at the end of the previous
step.
{\bf 6.} Go to step 2. 
After this loop is repeated many times,
 we generate a histogram of $x$. 
 The histogram so created
is identical to $\langle P(x,t) \rangle$ when $t$ is large.
On a computer the second step is implemented with a simple
algorithm provided by Chambers et al \cite{Chambers}.
Notice that with this exact scheme 
we have mapped the random walk in a random environment
to a Brownian motion problem.  
We see that for quenched disorder
 the operational
time is ${\cal S}_\alpha$ and in this sense subordination is weakly
broken: the L\'evy transformation \cite{Sokolov} is still maintained.  

\begin{figure}
\begin{center}
\epsfig{figure=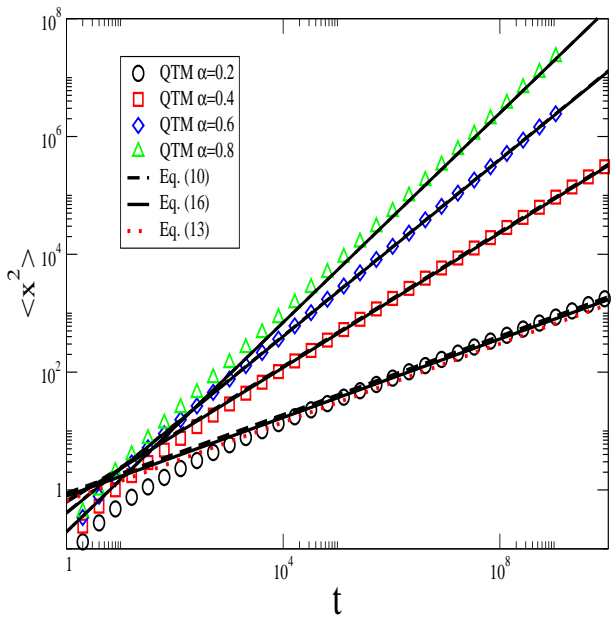,totalheight=0.34\textheight,
width=0.45\textwidth,angle=-90}
\end{center}
\caption{ The mean square displacement of the QTM
versus time. Numerical data matches perfectly the theory based on 
weak subordination breaking (the lines plotted with $\langle z^2 \rangle$ in
Table \ref{table1}) and analytical formulas
Eq. (\ref{eqaplz13}) for $\alpha=0.2$ and 
Eq. 
(\ref{eqBa204})
for $\alpha=0.4,0.6,0.8$. 
}
\label{fig2}
\end{figure}

{\em The diffusion front of the QTM.}
 Let $P_{{\cal S}_\alpha} (x)$ be the PDF of $x$ for the simple
random walk on a lattice (Brownian motion) stopped at the
 operational time
${\cal S}_\alpha$. 
Since the QTM dynamics can be separated into  two distinct
processes:  Brownian motion with operational time ${\cal S}_\alpha$ 
(step {\bf 4.})  and
the L\'evy time transformation (steps {\bf 2.} and {\bf 3.})
we find
\begin{equation}
\langle P(x,t) \rangle \sim \int_0 ^\infty P_{{\cal S}_\alpha} (x)  
n_{t} \left( {\cal S}_\alpha\right) {\rm d} {\cal S}_\alpha
\label{eqP01} 
\end{equation} 
where $n_t({\cal S}_\alpha)$ is given in Eq. (\ref{eqtime05}).
For the mean field version of the model (i.e. CTRW) replace ${\cal S}_\alpha$ 
with the number of steps $s$ of the Brownian motion, and then 
$P_{{\cal S}_\alpha } (x)$ is 
Gaussian as well known \cite{Sokolov}. 
 From  normal Brownian  motion
we have the scaling behavior  
 $x \propto ({\cal S}_\alpha)^{1/(1 +\alpha)}$.
To see this we use: (i) usual Brownian scaling
$x \propto s^{1/2}$ (ii)
$n_x$
 within a region $|x|<s^{1/2}$  is roughly 
the number of jumps made $s$ divided by the number of sites 
in the explored
region 
$n_x \propto 
s/s^{1/2}= s^{1/2}$.
Hence ${\cal S}_\alpha \propto \sqrt{s} (n_x)^\alpha  \propto s^{(1+\alpha)/2}$
which gives $x \propto ({\cal S}_\alpha)^{1/(1 +\alpha)}$.
This scaling implies 
\begin{equation}
P_{{\cal S}_\alpha} (x) = {1 \over ({\cal S}_\alpha )^{1/(1+\alpha)}}
 B_\alpha\left[{ x \over ({\cal S}_\alpha)^{1/(1 +\alpha)} } \right]   
\label{eqSca03}
\end{equation} 
with $B_\alpha(z)$ a normalized non negative function.
Define the 
scaling variable
%
$ \xi =  x / \left( t/A^{1/\alpha}  \right)^{{\alpha \over 1 + \alpha}}$
%
and 
%
$\langle P(x,t) \rangle  \sim   g_\alpha \left( \xi \right)/ \left( t/A^{1/\alpha} \right)^{{ \alpha \over 1 + \alpha }} $
%
which according to Eq. (\ref{eqP01}) is
\begin{equation}
g_\alpha\left(\xi\right) = \int_0 ^\infty {\rm d} y y^{ { \alpha \over 1 + \alpha }} B_\alpha \left( \xi y^{{ \alpha \over 1 + \alpha }} \right) l_{\alpha,1,1}\left( y \right) . 
\label{eqSca06}
\end{equation} 
%
 A general relation is found between the moments 
$\langle |x|^q \rangle = \langle 
\int_{-\infty} ^\infty |x|^q P(x,t) {\rm d} x \rangle
$ of the original QTM and the moments 
$\langle |z|^q \rangle=\int_{-\infty} ^\infty |z|^q B_{\alpha} (z) {\rm d}z$
\begin{equation}
\langle |x|^q \rangle = \langle |z|^q \rangle { \Gamma\left( { q \over 1 + \alpha} \right) \over \alpha \Gamma \left( { q \alpha \over 1 + \alpha } \right) } \left( {t \over A^{1/\alpha} } \right)^{{\alpha q \over 1 + \alpha} } .
\label{eqSca08}
\end{equation} 
The new content of Eqs.
(\ref{eqSca06},
\ref{eqSca08}) is that once we obtain $B_\alpha(z)$ 
either from theory or simulations of Brownian trajectories,
we have a useful method to obtain
exact  statistical properties of the diffusion front.

\begin{figure}
\begin{center}
\epsfig{figure=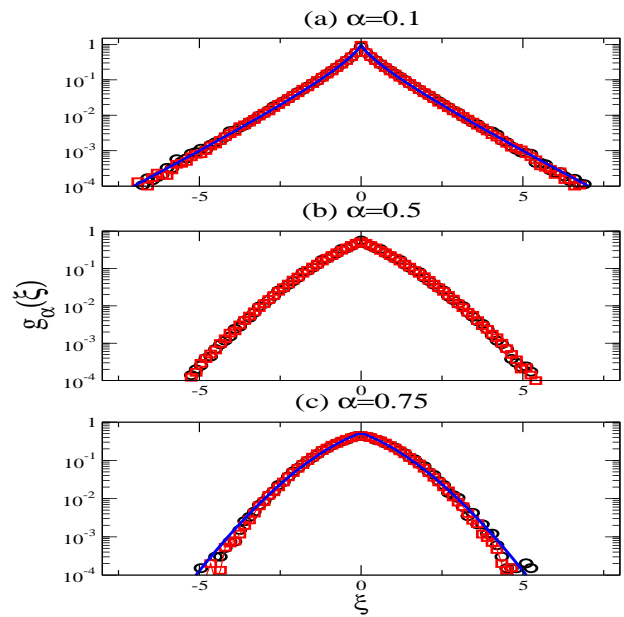,totalheight=0.34\textheight, width=0.45\textwidth}
\end{center}
\caption{ The diffusion front  of QTM (squares) perfectly
matches  theory based on weak sub-ordination breaking (circles)
and analytical predictions (lines)  Eqs. 
(\ref{eqSca06},\ref{eqaplz12},\ref{eqBa102a}).  
}
\label{fig3}
\end{figure}

 Generating Brownian trajectories on a lattice we found
$B_\alpha(z)$ in Fig. \ref{fig1}, which shows an
interesting transition from a $V$ shape when $\alpha\to 0$ to a Gaussian shape,
which we soon analyze analytically. 
With $\langle z^2 \rangle$ given  in Table \ref{table1} 
and Eq. (\ref{eqSca08}) we get the mean square displacement of the
QTM $\langle x^2 \rangle$. We then
favorably  compare 
the predictions of our theory with simulations
of the QTM in Fig. \ref{fig2} (and analytical formulas soon developed). 
In Fig. \ref{fig3}
we show $g_\alpha(\xi)$ and present 
excellent agreement between weak subordination breaking 
and direct simulation of the QTM. 
One advantage of our approach is 
that it is capable of dealing with the critical slowing down
pointed out by Bertin and Bouchaud \cite{Bertin}. Briefly,
QTM simulations do not converge on reasonable
computer time scales for say $\alpha>0.8$. In contrast 
weak subordination breaking scheme quickly converges 
since it is based on Brownian motion and there is no need
to generate disordered systems. More importantly we now
analyze Brownian motion analytically, obtain $B_\alpha(z)$ in two
important limits and then with Eqs. 
(\ref{eqSca06},
\ref{eqSca08})
provide solutions
to the QTM.  

{\em The limit $\alpha \to 0$} corresponds to strong disorder.
To find $B_0(z)$  we consider
Brownian motion stopped at 
 ``time"
 ${\cal S}_0$ where as
mentioned ${\cal S}_0$ is the span
of the random walk. 
Consider
$P_{{\cal S}_0} \left( {\cal S}_0 -n\right)$ where $x={\cal S}_0 - n > 0$
and for simplicity we start with $n=1$. 
The Brownian
particle after the first step can
be either on $x=1$ or $x=-1$. If it is on $x=-1$ it must travel
a distance ${\cal S}_0$ to reach its destination $x={\cal S}_0 - 1$ and
the span is ${\cal S}_0$. On the other hand
if it jumps to  $x=1$ 
the distance the particle must travel is ${\cal S}_0 - 2$ and
the span must still be ${\cal S}_0$. Hence 
%
$P_{{\cal S}_0} \left( {\cal S}_0 - 1 \right) = [ P_{{\cal S}_0 }\left( {\cal S}_0\right)+ P_{{\cal S}_0} \left( {\cal S}_0 - 2  \right) ] /2.$
%
More generally
\begin{equation}
 P_{{\cal S}_0 } \left({\cal S}_0 -n \right) =
{ 1 \over 2} \left[ P_{{\cal S}_0 } \left({\cal S}_0 -n-1 \right) +
 P_{{\cal S}_0} \left({\cal S}_0 -n+1  \right)\right], 
\label{eqalpz03}
\end{equation}  
and for the boundary term
$ P_{{\cal S}_0} ({\cal S}_0) =
 \left[  P_{{\cal S}_0} \left( {\cal S}_0 - 1 \right) +  P_{{\cal S}_0 -1} \left( {\cal S}_0 - 1 \right) \right]/2.$
Eq. (\ref{eqalpz03}) is easily solved
$P_{{\cal S}_0 } \left(x \right) = { |x| \over {\cal S}_0 \left( {\cal S}_0 + 1 \right) }$  for $-{\cal S}_0   \le x \le {\cal S}_0 $ 
and $x \in {\bf Z}$. 
In the limit ${\cal S}_0\gg 1$ we have for the scaled variable
 $z=x/{\cal S}_0$ the $V$  shape PDF (see Fig. \ref{fig1}) 
\begin{equation} 
\lim_{\alpha \to 0} B_{\alpha} (z) = \left\{
\begin{array}{l l}
|z| & \ \mbox{for} \ |z|<1 \\
\  & \  \\
0 & \mbox{otherwise} .
\end{array}
\right.
\label{eqaplz05}
\end{equation} 
This $V$ shape reflects the tendency of a Brownian
 particle to reach a large span ${\cal S}_0$
when it is far from the origin. 

According to 
Eq. (\ref{eqSca08})
the even moments 
$\langle x^{2q} \rangle$ for the random walk in
the QTM are given once we obtain
 $\langle z^{2q} \rangle$.
 In the limit $\alpha \to 0$ we
find using Eq. 
(\ref{eqaplz05})
$\langle z^{2 q} \rangle = 2 \int_0 ^1 z^{2 q}  z {\rm d} z = (1 + q)^{-1}$
hence with Eq. 
(\ref{eqSca08})
we have for small $\alpha$
\begin{equation}
\langle x^2 \rangle \simeq { 1 \over 2} { \Gamma\left( { 2 \over 1 + \alpha} \right) \over \alpha \Gamma\left( {2  \alpha  \over 1 + \alpha} \right)} \left(
{ t \over A^{1 /\alpha} } \right)^{2 \alpha  \over 1 + \alpha} 
\label{eqaplz13}
\end{equation}
which is tested in Fig. \ref{fig2}. 
Inserting $\langle z^{2q}\rangle=(1+q)^{-1}$ 
in Eq. 
(\ref{eqSca08})
 we obtain the moments $\langle x^{2 q}\rangle$  of the QTM.  
Straight forward analysis then gives
\begin{equation} 
\lim_{\alpha \to 0} g_\alpha(\xi) = e^{-|\xi|}-|\xi| E_1 \left( |\xi|\right) 
\label{eqaplz12} 
\end{equation} 
where $E_1(\xi) = \int_\xi ^\infty (e^{- t} / t){\rm d} t$ is the tabulated
exponential integral. This scaling function
was obtained by C. Monthus  \cite{Monthus} using an 
RG method which is exact
in the limit $\alpha \to 0$.

{\em Approaching the weak disorder limit $\alpha \to 1$.}
For $\alpha=1$ we have ${\cal S}_1 = \sum_{x=-\infty} ^\infty n_x= s$, namely
${\cal S}_1$ is {\em non}  random since it 
is equal to the number of steps made. 
Therefore  when
$\alpha$ is close enough to $1$  we
may neglect fluctuations
and 
 ${\cal S}_\alpha =\langle {\cal  S}_\alpha \rangle$.
In a longer publication we will show
that 
$\langle {\cal S}_\alpha \rangle=  C_\alpha s^{{1 + \alpha \over 2 }}$. 
and ${\cal C}_\alpha =  2^{(\alpha + 3)/2 } \Gamma\left( 1 + \alpha / 2\right) /[ \sqrt{\pi} \left( 1 + \alpha \right)]$
[hints: find $\langle (n_x)^\alpha\rangle$ with  the
first passage time PDF for Brownian motion and renewal theory, 
for the limit $\alpha\to 0$ consult \cite{Erdos}].
As well known the PDF of finding the Brownian particle on
$x$ after $s$ jumps is the Gaussian
%
$P_s(x) =  \exp\left( - x^2 / 2 s \right) / \sqrt{ 2 \pi s} $
%
hence change of variables to ${\cal S}_\alpha$ gives 
\begin{equation}
B_\alpha(z) \sim { \exp\left[ - { \left( C_\alpha \right)^{{ 2 \over 1 + \alpha}} z^2  \over 2} \right] \over \left[ 2 \pi / (C_\alpha)^{ { 2 \over 1 + \alpha}} \right]^{1/2} }.
\label{eqBa102a} 
\end{equation} 
It follows that 
$\langle z^2 \rangle\sim \left( C_\alpha \right)^{-{ 2 \over 1 + \alpha }}$
hence for the QTM 
\begin{equation}
\langle x^2 \rangle \simeq \left( C_\alpha\right)^{-{ 2 \over 1 + \alpha }} { \Gamma\left( { 2 \over 1 + \alpha} \right) \over \alpha \Gamma\left( { 2 \alpha \over 1 + \alpha } \right) } \left( { t \over A^{1/\alpha} } \right)^{{ 2 \alpha \over 1 + \alpha}} .
\label{eqBa204}
\end{equation}
In Fig. \ref{fig1} $B_\alpha(z)$ obtained
from Brownian simulations is favorably compared with 
Eq. (\ref{eqBa102a}) for $\alpha=0.9$. Surprisingly, as we show
in Fig. \ref{fig2}
Eq. (\ref{eqBa204}) works very well even for $\alpha=0.4$.
With Eqs. 
(\ref{eqSca06},
\ref{eqBa102a})
 and steepest descent method we find
for  $\xi>>1$
\begin{equation}
g_\alpha \left( \xi \right) \sim b_1 \xi^{ - 2 { 1 - \alpha \over 3 - \alpha }} 
e^ { - b_2 \xi^{2 { 1 + \alpha \over 3 - \alpha }}}
\label{eqBa208}
\end{equation}
with
$b_1=\sqrt{(1+\alpha)/[2\pi \alpha (3-\alpha)]}D,
b_2=\left[(3-2\alpha)/2\right]D^2\ \mbox{and} \
D=\left[(1+\alpha)^{1-\alpha}\alpha^\alpha\cal{C}_\alpha\right]^{1/(3-\alpha)}$
which approach the expected normal Gaussian limit when $\alpha \to 1$.
In the opposite limit $\xi<<1$ 
$g_\alpha \left( \xi \right) \sim
 1/\sqrt{2 \pi}  - 2^{(\alpha -1)/2 } \left[ 
 (1 + \alpha)/ \alpha\right] \left\{ C_\alpha / \Gamma\left[ ( 1 - \alpha)/2 
\right] \right\} \xi^\alpha + \cdots$.

 The method presented here is not limited to one dimension neither
to unbiased motion.  For example
consider  the QTM  on a regular lattice in three dimensions.  
From ordinary Brownian motion we expect ${\cal S}_\alpha \sim c_\alpha s$, where
now ${\cal S}_\alpha$ is non random and the (non trivial)
parameter $c_\alpha$ will depend
on the lattice structure. In principle once
$c_\alpha$ is determined one can map the
QTM with an equation like Eq.
(\ref{eqP01}) to an ordinary Brownian motion. This 
will yield mean field
CTRW dynamics which is still non trivial since the transformation
depends on the parameter  $c_\alpha$. Note that the QTM dynamics
describes also 
certain models of random walks on random geometries, e.g. fractal
comb structures and  naturally our approach captures also these cases.    

 Finally, 
weak subordination breaking scheme is not only a new
approach which deals with anomalous diffusion 
in systems with quenched disorder. 
Our method can be used to solve 
other aspects of dynamics in disordered systems like
aging and weak ergodicity breaking. 

{\bf Acknowledgment} This work was supported by the Israel science foundation.
We thank S. Majumdar for correspondence and Z. Shemer for many discussions.


\begin{thebibliography}{99}

\bibitem{Bouchaud} J. P. Bouchaud, and A. Georges, {\em Phys. Rep.}
{\bf 195}, 127 (1990).

\bibitem{Review} R. Metzler, J. Klafter, {\em Phys. Rep.} {\bf 339}, 1 (2000).

\bibitem{Scher} H. Scher, E. W. Montroll {\em Phys. Rev. B} {\bf 12} 
2455 (1975). 

\bibitem{B92} J. P. Bouchaud, {\em J. Phys. } I France {\bf 2}, 1705 (1992). 

\bibitem{MB96}
 C. Monthus and J.-P. Bouchaud, 
{\em J. Phys. A} {\bf  29},  
3847
(1996).

\bibitem{Stas} 
S. Burov, E. Barkai, 
{\em Phys. Rev. Lett.} {\bf 98} 250601 (2007).

\bibitem{Golan} 
G. Bel, E. Barkai
{\em Phys. Rev. Lett.} {\bf 94} 240602 (2005).

\bibitem{PhysToday}  
F. D. Stefani, J. P. Hoogenboom, and E. Barkai
{\em Physics Today} {\bf 62} nu. 2,  p. 34 (February 2009).

\bibitem{Burov} 
Y. He, S. Burov, R. Metzler, E. Barkai
{\em Phys. Rev. Lett.} {\bf 101}, 058101 (2008).

\bibitem{Machta} J. Machta, 
{\em  J. Phys. A: Math. Gen.} 
{\bf 18} no. 9, L531 (1985). 

\bibitem{Alexander1} S. Alexander {\em Phys. Rev. B} {\bf 23} 2951 (1981). 

\bibitem{Ben} G.  Ben Arous and J. $\breve{\mbox{C}}$ern\'y
{\em Ann. Probab.} {\bf 35} 2356 (2007). 

\bibitem{BenHarosh} G.  Ben Arous and J. $\breve{\mbox{C}}$ern\'y 
Lecture Notes for Les Houches Summer School Mathematical statistical physics, arXiv:math/0603344 (2008). 


\bibitem{Monthus} C. Monthus 
{\em Phys. Rev. E} {\bf 68} 036114 (2003). 

\bibitem{Sokolov} 
H. C. Fogedby {\em Phys. Rev. E.} {\bf 50} 1657 (1994).
E. Barkai
{\em Phys. Rev. E} {\bf 63}, 046118 (2001).
I. M. Sokolov,  and J. Klafter {\em Chaos} {\bf 15}, 026103 (2005). 
E. Heinsalu, M. Patriarca, I. Goychuk, P. H\"anggi {\em Phys. Rev. Lett.} 
{\bf 99} 120602 (2007). 
 M. Magdziarz, A. Weron, and J. Klafter 
{\em Phys. Rev. Lett.}  {\bf 101}, 210601 (2008). 
S. Eule, R. Friedrich 
{\em Europhysics Letters} {\bf  86}, 30008 (2009).
B. I. Henry, T. A. M. Langlands, P. Straka {\em Phys. Rev. Lett.} {\bf 105}, 
170602 (2010). 

\bibitem{Bertin} E. M. Bertin and J. P. Bouchaud, {\em Phys. Rev. E.} {\bf 67} 026128 (2003). 



\bibitem{Chambers} J. M. Chambers, C. L. Mallows and B. W. Stuck
{\em J. of the American Statistical Association} {\bf 71} (354) 340 (1976). 

\bibitem{Erdos} A. Dvoretzky and E. Erd$\ddot{\mbox{o}}$s in Proc. 2nd Berkeley Symp.
Math. Stat. and Prob. (University of California Press) 33 (1951). 





\end{thebibliography}
\end{document}